\documentclass[12pt,a4paper]{article}
\usepackage[english]{babel}
\usepackage{graphicx}

\newcommand{\alt}{\mathbin{\lower 3pt\hbox
   {$\rlap{\raise 5pt\hbox{$\char'074$}}\mathchar"7218$}}}
\newcommand{\agt}{\mathbin{\lower 3pt\hbox
   {$\rlap{\raise 5pt\hbox{$\char'076$}}\mathchar"7218$}}}

\begin{document}
\setcounter{footnote}{0}
\setcounter{equation}{0}
\setcounter{figure}{0}
\setcounter{table}{0}

\begin{center}
{\large\bf Quantum Electrodynamics at Extremely Small Distances}

\vspace{4mm}
I. M. Suslov \\
P.L.Kapitza Institute for Physical Problems,
\\ 119337 Moscow, Russia \\
\vspace{5mm}
\end{center}

\begin{center}
\begin{minipage}{135mm}
{\bf Abstract} \\
The asymptotics of the Gell-Mann -- Low function in QED can be
determined exactly, $\beta(g)= g$ at $g\to\infty$, where $g=e^2$
is the running fine structure constant. It solves the problem of
pure QED at small distances $L$ and gives the behavior $g\sim
L^{-2}$.

\end{minipage}
\end{center}
\vspace{5mm}


According to Landau, Abrikosov, Khalatnikov \cite{100}, relation
of the bare charge $e_0$ with the observable charge $e$ in quantum
electrodynamics (QED) is given by expression
$$
e^2=\frac{e^2_0}{1+\beta_2 e^2_0 \ln \Lambda^2/m^2}  \,,
\eqno(1)
$$
where $m$ is the mass of the particle, and $\Lambda$ is
the momentum cut-off.  For finite $e_0$ and
$\Lambda\to \infty$  the "zero charge"
situation ($e\to 0$) takes place. The proper interpretation of
Eq.1 consists in its inverting, so that $e_0$  (related to the
length scale $\Lambda^{-1}$)
is chosen to give a correct value of $e$:
$$
e^2_0=\frac{e^2}{1-\beta_2 e^2 \ln \Lambda^2/m^2} \,.
\eqno(2)
$$
The growth of $e_0$ with  $\Lambda$ invalidates Eqs.1,2
in the region $e_0\sim 1$
and existence of "the Landau pole" in Eq.2 has no physical sense.

The actual behavior of the charge  $e$ as a function of the
length  scale $L$ is determined by the Gell-Mann -- Low equation
\cite{101}
$$
-\frac{dg}{d \ln L^2} =\beta(g)=\beta_2 g^2+\beta_3 g^3+\ldots
\,,\qquad g=e^2\,,
\eqno(3)
$$
and depends on appearance of the function  $\beta(g)$. According
to classification by Bogolyubov and Shirkov \cite{102}, the growth
of  $g(L)$ is saturated, if $\beta(g)$ has a zero for finite  $g$,
and continues to infinity, if $\beta(g)$ is non-alternating and
behaves as $\beta(g)\sim g^\alpha$ with $\alpha\le 1$ for large
$g$; if, however, $\beta(g)\sim g^\alpha$ with $\alpha>1$, then
$g(L) $ is divergent  at finite $L=L_0$  (the real Landau pole
arises) and the theory is internally inconsistent due to
indeterminacy of $g(L)$ for $L<L_0$.
Landau and Pomeranchuk \cite{103} tried to justify the latter
possibility, arguing that Eq.1 is valid without restrictions;
however, it is possible only for the strict equality
$\beta(g)=\beta_2 g^2$, which is surely invalid due to finiteness
of $\beta_3$.
One can see that solution of
the  problem of QED at small distances
needs calculation of the Gell-Mann -- Low function $\beta(g)$ at
arbitrary $g$, and in particular its asymptotic behavior for
$g\to\infty$.

It was found recently in \cite{1}, that strong coupling
asymptotic behavior of the renormalization group functions in
the actual field theories can be obtained analytically.
Attempts to reconstruct  the $\beta$-function for
$\varphi^4$ theory, undertaken  previously by summation of
perturbation series, lead to asymptotics $\beta(g)=\beta_\infty
g^\alpha$ at $g\to\infty$, where $\alpha\approx 1$ for space
dimensions $d=2,3,4$  \cite{4,14,15}. The natural
hypothesis arises, that the true
asymptotic behavior is $\beta(g) \sim g$ for all
$d$.  Consideration of the "toy" zero-dimensional model
confirms the hypothesis and reveals the origin of the linear
asymptotics. It is related with unexpected circumstance that the
 strong coupling limit for the renormalized charge  $g$
 is determined not by large values of the bare
 charge  $g_0$,  but its complex values.
More than that, it is sufficient to consider the region $|g_0|\ll
1$, where the functional integrals can be evaluated in the
saddle-point approximation. If a proper direction in the complex
$g_0$ plane is chosen, the
contribution of the
trivial vacuum is comparable with the saddle-point contribution
of the main instanton, and a functional integral can turn to
zero. The limit $g\to\infty$ is related with a zero of a
certain functional integral and appears to be completely
controllable.  As a result, it is possible to obtain asymptotic
behavior of the $\beta$-function and anomalous dimensions: the
former indeed appears to be linear.

At the present paper we show that the same idea can be applied to
QED. Attempted reconstruction of the $\beta$-function in this
theory  \cite{5} gives for it non-alternating behavior (Fig.1)
\begin{figure}
\centerline{\includegraphics[width=5.1 in]{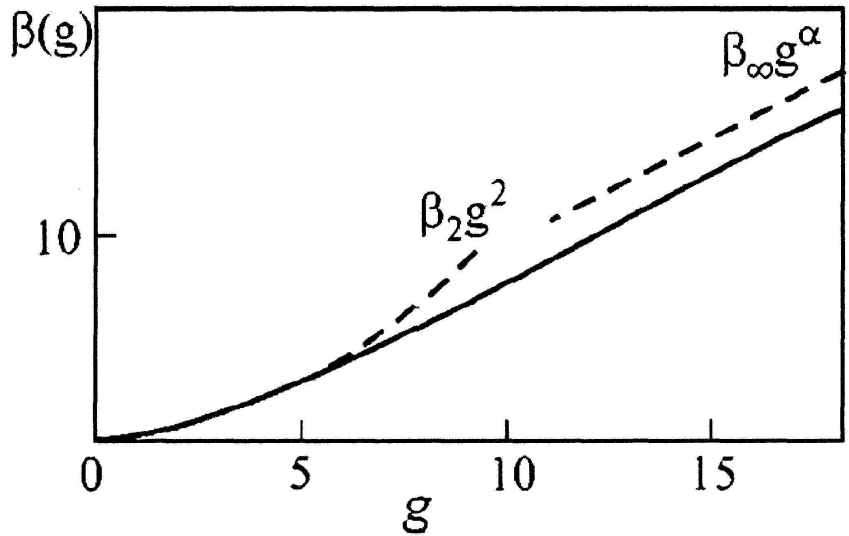}} \caption{
General appearance of the $\beta$-function in QED.} \label{fig1}
\end{figure}
with the asymptotics
$\beta(g)=\beta_\infty g^\alpha$, where
 $$
 \alpha=1.0\pm  0.1\,,\qquad \beta_\infty=1.0\pm 0.3 \,.
 \eqno(4)
 $$
 Within uncertainty, the obtained  $\beta$-function
satisfies inequality
 $$ 0\le \beta(g)< g \,,
 \eqno(5)
 $$
 established in \cite{6,7} from the spectral representations,
 while the asymptotics  (4) corresponds to the upper bound of (5).
 Such coincidence does not look incident and indicates that
 asymptotics $\beta(g)= g$ is an exact result.
 We show below that it is so indeed.

The general functional integral of QED contains  $M$
photon and $2N$  fermionic fields in the
pre-exponential,
  $$
  I_{M,2N}=\int DA D\bar\psi D\psi\, A_{\mu_1}(x_1)\ldots
  A_{\mu_M}(x_M)\,
          \psi(y_1)\bar\psi(z_1)\ldots \psi(y_N)\bar\psi(z_N)
\exp\left(-S\{A,\psi,\bar\psi\}  \right) \,,
  \eqno(6)
  $$
where $S\{A,\psi,\bar\psi\}$ is the Euclidean action,
$$
S\{A,\psi,\bar\psi\} =
\int d^4x \left[ \frac{1}{4}
               (\partial_\mu A_\nu- \partial_\nu A_\mu)^2
        +\bar\psi(i\!\!\not{\! \partial} -m_0 +
    e_0 \! \not  {\!\! A})\psi \right]\,,
  \eqno(7)
$$
while $e_0$ and $m_0$ are the bare charge and mass.
Fourier transforms of the integrals $I_{M,N}$ with excluded
$\delta$-functions of the momentum conservation will
be referred as $K_{MN}(q_i,p_i)$ after extraction
of the usual
factors depending on tensor indices\,\footnote{\,A specific form
of these factors is inessential, since the results are
independent on the absolute normalization of $e$ and $m$.}; $q_i$
and $p_i$ are momenta of photons and electrons. Introducing
Green's functions $G^{(M,N)}=K_{MN}/K_{00}$, we can define
amputated vertices $\Gamma^{(M,N)}$ with $M$ photon and $N$
electron external legs:
$$
\Gamma^{(0,2)}(p)=1/G^{(0,2)}(p)\equiv 1/G(p)\,,\qquad
\Gamma^{(2,0)}(q)=1/G^{(2,0)}(q)\equiv 1/D(q)\,,\qquad
$$
$$
G^{(1,2)}(q,p,p')=D(q) G(p) G(p')\Gamma^{(1,2)}(q,p,p')\,,\qquad
  \eqno(8)
$$
etc., where $G(p)$ and $D(q)$ are the exact electron and photon
propagators.

Multiplicative renormalizability of the vertex  $\Gamma^{(M,N)}$
means that
$$
\Gamma^{(M,N)}(q_i,p_i; e_0, m_0, \Lambda) = Z_3^{-M/2}
Z_2^{-N/2} \Gamma_R^{(M,N)}(q_i, p_i; e,m)\,,
\eqno(9)
$$
i.e. its divergency at $\Lambda\to \infty$ disappears
 after extracting the proper
$Z$-factors and transferring to the renormalized charge $e$ and
mass $m$.  Renormalization conditions at zero momenta  are
accepted
$$
\left.\Gamma_R^{(0,2)}(p)\right|_{p\to 0} =\,\not{\!p} - m
\,, \qquad
$$
$$
\left.\Gamma_R^{(2,0)}(q)\right|_{q\to 0} =q^2\,, \eqno(10)
$$
$$
\left.\Gamma_R^{(1,2)}(q,p,p')\right|_{q,p,p'\to 0} = e\,,
$$
where the usual pole structure of the electron and photon
propagators is taken into account.  Substitution
of (10) into  (9) determines  $e$, $m$, $Z_2$,
$Z_3$ in terms of the bare quantities
$$
Z_2 = \left( \frac{\partial}{\partial\!\! \not{\! p}}
\left. \Gamma^{(0,2)}(p; e_0, m_0, \Lambda) \right|_{p=0}
\right)^{-1}
$$
$$
Z_3 = \left( \frac{\partial}{\partial q^2}
\left. \Gamma^{(2,0)}(q; e_0, m_0, \Lambda) \right|_{ q = 0}
\right)^{-1}
\eqno(11)
$$
$$
m=-Z_2
\left. \Gamma^{(0,2)}(p; g_0, m_0, \Lambda) \right|_{p=0}
$$
$$
e= Z_2 Z_3^{1/2}
\left. \Gamma^{(1,2)}(q,p,p'; e_0, m_0, \Lambda)
\right|_{q,p,p'=0}
$$
The Gell-Mann -- Low function is defined in this scheme  as
$$
\beta(g)=\left. \frac{d g}{d \ln m^2} \right|_{e_0,\Lambda=const}
\,,\qquad
\eqno(12)
$$
where $g=e^2$ is the running fine structure constant.
Using Eq.8 and definition of $G^{(M,N)}$, one has
$$
\Gamma^{(0,2)}(p)= \frac{K_{00}}{K_{02}(p)}\,,\qquad
\Gamma^{(2,0)}(q)= \frac{K_{00}}{K_{20}(q)}\,,\qquad
\Gamma^{(1,2)}= \frac{K_{12} K_{00}^2}{K_{02}^2 K_{20}}\,,\qquad
\eqno(13)
$$
where zero momenta are implied in the last relation.
Setting for small momenta
 $$
K_{02}(p)=K_{02}+\tilde K_{02} \!\!\not{\!p}\,,\qquad
K_{20}(q)=K_{20}+\tilde K_{20} q^2\,,\qquad
\eqno(14)
$$
and taking (11) into account, we have
$$
Z_2=- \frac{K_{02}^2}{K_{00} \tilde K_{02}}\,,\qquad
Z_3=- \frac{K_{20}^2}{K_{00} \tilde K_{20}}\,,\qquad
m= \frac{K_{02}}{ \tilde K_{02}}\,,\qquad
g=- \frac{K_{12}^2 K_{00}}{ \tilde K_{02}^2 \tilde
K_{20}}\,,\qquad \eqno(15)
$$
Denoting  differentiation over $m_0$ by prime, one has from (15)
$$
\frac{d m}{d m_0}=\left( \frac{K_{02}}{ \tilde K_{02}} \right)'
=\frac{  K'_{02}\tilde K_{02} - K_{02}\tilde K'_{02}}
{ \tilde K_{02}^2}
\eqno(16)
$$
Since  differentiation in (12) occurs at
$e_0,\,\,\Lambda = const$, the latter parameters are considered
to be fixed throughout all calculations: then $m$ is a
function of only $m_0$ and Eq.16 defines also the derivative
$d m_0/d m$. Using  definition of
the $\beta$-function (12),
$$
\beta(g)= \frac{m}{2}
\left(-\frac{K_{12}^2 K_{00}}{ \tilde K_{02}^2 \tilde K_{20}}
 \right)'_{m_0} \,
\frac{d m_0}{d m} \,,
\eqno(17)
$$
and making simple transformations, we end with equations
$$
g= -\frac{K_{12}^2 K_{00}}{ \tilde K_{02}^2 \tilde K_{20}}\,,
\qquad
  \eqno(18)
$$
$$
\beta(g)= \frac{1}{2} \frac{K_{02} \tilde K_{02}}
{ K_{02}\tilde K'_{02} -K'_{02}\tilde K_{02} }
\frac{K_{12}^2 K_{00}}{ \tilde K_{02}^2 \tilde K_{20}}
\left\{ \frac{2\,K'_{12}}{ K_{12}} +\frac{K'_{00}}{ K_{00}}
-\frac{2\,\tilde K'_{02}}{\tilde K_{02}}
-\frac{\tilde K'_{20}}{\tilde K_{20}}
\right\}
  \eqno(19)
$$
Equations (18),(19) define the dependence $\beta(g)$ in the
parametric form. Their right hand sides depend on $m_0$, $g_0$,
$\Lambda$ with two latter parameters being fixed.
Solving Eq.18 for $m_0$  and substituting into (19),
one obtains  $\beta$ as a function of $g$, $g_0$ and
$\Lambda$; in fact, the dependence on
the latter two parameters is absent due to the general theorems
\cite{16,17}.

According to  \cite{1}, the strong coupling regime for
renormalized interaction is related with a zero of a certain
functional integral. It is clear from (18) that the limit
$g\to\infty$ can be realized by two ways:  tending to zero either
$\tilde K_{02}$, or $\tilde K_{20}$.  For $\tilde K_{02}\to 0$,
equations (18,19) are simplified,
$$
g= -\frac{K_{12}^2 K_{00}}{ \tilde K_{02}^2 \tilde K_{20}}\,,
\qquad
\beta(g)= -\frac{K_{12}^2 K_{00}}{ \tilde K_{02}^2 \tilde
K_{20}}\,,
\eqno(20)
$$
and the parametric representation is resolved in the form
$$
\beta(g)= g\,,\qquad g\to \infty\,.
\eqno(21)
$$
For $\tilde K_{20}\to 0$, one has
$$
g\propto \frac{1}{\tilde K_{20}}\,,\qquad
\beta(g)\propto \frac{1}{\tilde K^2_{20}}\,,
\eqno(22)
$$
and hence
$$
\beta(g)\propto g^2\,,\qquad g\to \infty\,.
\eqno(23)
$$
Consequently, there are two possibilities for the
asymptotics of  $\beta(g)$, either (21), or (23). The
second possibility is in conflict with inequality (5),
while the first possibility is in excellent agreement with
results (4) obtained by summation of perturbation series.
In our opinion, it is sufficient reason to consider
 Eq.21 as an exact result for the asymptotics of the
 $\beta$-function. It means that the fine structure
 constant in pure QED behaves as  $g\propto L^{-2}$ at small
 distances $L$.

Above we have in mind that the mechanism determining
the asymptotics of  $\beta(g)$ is the same as in
$\varphi^4$ theory. Strictly speaking, one cannot exclude
possibility that the strong coupling regime in QED is
determined by the different mechanism, e.g.
by the large value  of the integral  $K_{12}$. However, such
possibility looks improbable: if one make rough estimate
of pre-exponential in (6), supposing that all fields are
localized at the unit length scale,
$$
K_{12}\sim \langle A \rangle \langle \psi \bar\psi \rangle
K_{00}\,, \qquad
\tilde K_{02}\sim
K_{02}\sim \langle \psi \bar\psi \rangle K_{00}\,, \qquad
\tilde K_{20}\sim
K_{20}\sim \langle A \rangle^2 K_{00}\,,
\eqno(24)
$$
then substitution into (18) gives $g\sim 1$. The change of the
general length scale  does not affect the quantity $g$ due to
its dimensionless character.  Consequently, it is
impossible to achieve large values of $g$  by a simple change
of the amplitude of  fields  $A$, $\psi$, $\bar\psi$,
or the scale of their spatial localization.
It is necessary to suggest that the average value of
$\langle A \rangle$ or $\langle \psi \bar\psi \rangle$
is anomalously small for one of the integrals, but it returns
us to already considered possibilities.

In the analogy with \cite{1}, the zeroes of functional integrals
can be obtained for the complex  $g_0$ with $|g_0|\ll 1$ by
compensation of the saddle-point contribution of  trivial
vacuum with the saddle-point contribution of the instanton
configuration with minimum action. The latter contribution is
well-known from the studies of large-order behavior of
perturbation series \cite{10,12,13,5} and has the form
$$
\left[K_{M,N}(q_i,p_i)\right]^{inst}=
i c(q_i,p_i) \left(\frac{S_0}{g_0^2} \right)^{b}
 {\rm e}^{-S_0/g^2_0}
\eqno(25)
$$
where  $S_0$ is the instanton action, $b=(M+r)/2$ and $r$ is
the number of zero modes. Setting  $t^2=-S_0/g^2_0$,
we come to expressions of the same kind, as were analyzed
in \cite{1}. It is easy to be convinced that zeroes of different
integrals  $K_{MN}$ and their derivatives lie in different
points.

In conclusion, we have determined the exact asymptotics of the
$\beta$-function in QED, which determines the behavior of the
effective interaction at extremely small distances.

\vspace{2mm}

This work is partially supported by RFBR  (grant 06-02-17541).

\end{document}